\documentclass[runningheads]{llncs}

\usepackage[T1]{fontenc}
\usepackage{graphicx}
\usepackage{amsmath,amssymb}
\usepackage{booktabs}
\usepackage{xspace}
\usepackage{listings}
\usepackage{url}
\usepackage{tikz}
\usetikzlibrary{arrows.meta,positioning,shapes}

\usepackage{xcolor}
\usepackage{listings}
\usepackage{comment}

\usepackage{tabularx}
\usepackage{array}
\usepackage{makecell}
\usepackage{hyperref}

\definecolor{why3keyword}{RGB}{0,70,140}
\definecolor{why3logic}{RGB}{120,40,120}
\definecolor{why3comment}{RGB}{90,110,90}
\definecolor{why3string}{RGB}{150,80,20}
\definecolor{why3background}{RGB}{248,248,248}

\definecolor{stageblue}{RGB}{232,241,252}
\definecolor{checkpurple}{RGB}{244,238,250}
\definecolor{testorange}{RGB}{255,246,230}
\definecolor{successgreen}{RGB}{232,246,236}
\definecolor{failred}{RGB}{252,235,235}
\definecolor{inputgray}{RGB}{247,247,247}

\lstdefinelanguage{Why3}{
  morekeywords={
    module,end,use,import,clone,export,scope,
    type,let,rec,function,predicate,val,lemma,axiom,goal,
    requires,ensures,invariant,variant,assert,assume,
    if,then,else,while,do,done,for,to,downto,
    match,with,try,raise,exception,any,abstract,
    forall,exists,true,false,old,result,ghost,ref,in
  },
  sensitive=true,
  morecomment=[l]{//},
  morecomment=[s]{(*}{*)},
  morestring=[b]",
}

\lstdefinestyle{why3style}{
  language=Why3,
  basicstyle=\ttfamily\small,
  keywordstyle=\color{why3keyword}\bfseries,
  commentstyle=\color{why3comment}\itshape,
  stringstyle=\color{why3string},
  backgroundcolor=\color{why3background},
  showstringspaces=false,
  columns=fullflexible,
  keepspaces=true,
  breaklines=true,
  frame=single,
  rulecolor=\color{black!20},
  frameround=tttt,
  numbers=left,
  numberstyle=\scriptsize\color{black!45},
  numbersep=8pt,
  xleftmargin=1.5em,
  framexleftmargin=1.2em,
  captionpos=b,
  literate=
    {->}{{$\to$}}2
    {<-}{{$\leftarrow$}}2
    {<=}{{$\leq$}}2
    {>=}{{$\geq$}}2
    {/\\}{{$\land$}}2
    {\\/}{{$\lor$}}2
    {<->}{{$\leftrightarrow$}}3
    {<>}{{$\neq$}}2
}

\newcommand{\stageA}{Stage~1\xspace}
\newcommand{\stageB}{Stage~2\xspace}
\newcommand{\stageC}{Stage~3\xspace}

\newcommand{\toolname}{{\sf Diversify2Verify}\xspace}
\newcommand{\whythree}{Why3\xspace}

\lstdefinelanguage{whyml}{
  keywords={module,use,import,predicate,function,let,rec,val,requires,ensures,variant,invariant,assert,forall,exists,if,then,else,match,with,end,true,false,ref,while,do,done,exception,raise,try,ghost},
  sensitive=true,
  comment=[l]{//},
  morecomment=[s]{(*}{*)},
  morestring=[b]"
}

\begin{document}

\title{Diversifying to Verify: When Task-Equivalent Programs Differ in Verifiability}
%
%
\author{Shirley Yu\inst{1} \and
Ruben Martins\inst{2}}
\authorrunning{Yu and Martins}
%
\institute{Stanford University, 
\email{shirleyyu0302@gmail.com}\\
\and
Carnegie Mellon University, 
\email{rubenm@andrew.cmu.edu}
}

\maketitle

\begin{abstract}
\begin{itemize}
Program verification is crucial for software correctness, but producing fully verified programs remains difficult in practice. This paper studies whether implementation structure affects automated verifiability when multiple generated programs are intended to satisfy the same task-level semantics.

We present \toolname, a staged LLM-based pipeline for \whythree that infers representation-specific contracts, generates and tests diverse recursive and imperative array/list implementations, and attempts  verification with bounded verifier-guided annotation repair.

We also construct a verification-oriented benchmark of 73 tasks over integers, arrays, and lists, yielding 292 implementation variants. \toolname verifies 96 artifacts initially and 154 after two repair passes, improving artifact-level verification from 32.9\% to 52.7\%. At the task level, at least one variant verifies for 49 of 73 tasks, a 67.1\% success rate. These results show that task-equivalent implementations can differ substantially in verifiability and that implementation diversity helps find verification-friendly artifacts.

\end{itemize}
\end{abstract}

\section{Introduction}

Large Language Models (LLMs) have become increasingly effective at synthesizing executable code from natural-language prompts~\cite{humaneval}. Generating formally verified software, however, requires more than code that passes tests. In deductive verification frameworks such as \whythree~\cite{filliatre2013why3}, a program must be paired with a formal semantic specification and proof guidance---including preconditions, postconditions, invariants, variants, assertions, helper predicates, and lemmas---sufficient for the verifier to prove correctness.
This makes verified-code generation difficult to treat as ordinary code generation. Natural-language tasks are often ambiguous or incomplete, and concrete examples validate only particular input-output pairs. A model must therefore infer a mathematical specification~\cite{postconditions,specgen}, implement the algorithm~\cite{humaneval,mbpp,alphacode}, and synthesize the proof structure needed to connect the implementation to that specification~\cite{dafny-ai,dafny-annotator}. In a one-shot approach, these requirements are entangled. When verification fails, feedback is often ambiguous, and unconstrained repair may change the specification rather than the implementation or proof, yielding a program that verifies for the wrong reason.

This paper asks whether implementation diversity can improve LLM-assisted deductive verification. \emph{Given multiple implementations intended to solve the same task, do different data representations and control structures lead to different verification outcomes?} Can generating several variants increase the chance that at least one has a verification-friendly structure? We use \emph{task-equivalent} to mean that variants are generated for the same benchmark task and verified against representation-specific contracts that are intended to express the same task-level semantics. We do not prove, as an additional theorem, that the array and list contracts are equivalent to each other.

We introduce \toolname, a staged LLM-based pipeline for generating and verifying \whythree artifacts from programming tasks and examples. \toolname separates contract inference, implementation generation, and proof annotation. It first generates a representation-specific contract with helper definitions and test lemmas derived from the examples, then freezes the accepted contract as the semantic target. It next generates executable WhyML candidates for the same task, varying representation and implementation style. Finally, it combines each tested implementation with the frozen contract and adds the annotations needed for \whythree verification. 

To evaluate this approach, we construct a verification-oriented benchmark from LeetCodeDataset~\cite{xia2025leetcodedataset}. Unlike standard code-generation benchmarks that primarily test executable behavior, our benchmark is designed to study deductive verification and implementation diversity. We restrict attention to tasks over integers, arrays, and lists that admit formal contracts, support multiple plausible implementations, and are suitable for SMT-backed reasoning. 
The resulting benchmark contains 73 tasks. For each task, \toolname generates four variants, array-recursive, array-imperative, list-recursive, and list-imperative, yielding 292 final verification artifacts.

Our evaluation shows that implementation structure substantially affects automated verifiability. Of the 292 generated artifacts, 96 verify initially and 154 verify after two bounded, feedback-guided repair passes, yielding an artifact-level verification rate of 52.7\%. At the task level, at least one variant verifies for 49 of the 73 tasks, for a success rate of 67.1\%. The strongest individual family verifies 44 tasks, indicating that diversity provides a modest but measurable increase in task coverage. Overall, the results support our central hypothesis: \emph{task-equivalent implementations can differ in verifiability}.

This paper makes the following contributions:
\begin{itemize}
\item \textbf{A staged LLM pipeline for \whythree.}
We present \toolname, which decomposes verified-code generation into contract inference, implementation generation, and proof annotation.

\item \textbf{Diverse implementation generation.}
For each task, \toolname produces recursive and imperative variants over array and list representations, enabling direct comparison of structurally different programs.

\item \textbf{A verification-oriented benchmark.}
We construct a benchmark of 73 tasks over integers, arrays, and lists, with accepted representation-specific contracts and 292 generated implementation artifacts. We release the benchmark, generated Why3 artifacts, and verification logs in a public repository available at \url{https://github.com/forge-lab/Diversify2Verify}.

\item \textbf{An empirical study of verifiability.}
We evaluate how verification success varies by representation and implementation style, quantify task-level gains from multiple variants, and analyze common remaining proof failures.
\end{itemize}

\section{Background: \whythree and Verification Challenges}
\label{sec:background}

This section reviews the \whythree concepts used by \toolname and explains how representation and control structure shape proof obligations, causing task-equivalent implementations to differ in verifiability.

\subsection{Deductive Verification in \whythree}
Deductive verification provides mathematical guarantees about software correctness by proving that an implementation satisfies a formal specification. In \whythree, programs are written in WhyML, which supports both functional and imperative programming. \whythree generates verification conditions from annotated programs and delegates them to external SMT solvers and automated theorem provers.

Correctness is expressed using contracts. A function's expected input state is specified with \texttt{requires} clauses, or preconditions, while its guaranteed output state is specified with \texttt{ensures} clauses, or postconditions. Iterative and recursive constructs usually require additional proof guidance. Loops use \texttt{invariant} clauses to state properties that hold before the loop and are preserved by every iteration. They also use \texttt{variant} clauses to provide a strictly decreasing measure, such as an integer bound or data-structure size, that proves termination. Recursive functions similarly require contracts and termination arguments.

\whythree also supports ghost code: proof-only variables, expressions, and statements that help verification without changing executable behavior. Ghost code is useful for recording logical facts that are implicit in the implementation, such as which part of the input has been processed, how an accumulator relates to the specification, or which witness supports a postcondition.

\subsection{Representation and Control Structure}

The choice of data representation affects the shape of the proof. Array programs typically require index-based reasoning. Every array access creates bounds obligations, and functional properties often require quantified facts over
ranges of indices. Imperative array algorithms therefore need loop invariants that summarize what is known about processed prefixes, suffixes, or windows.

List programs shift the burden from index arithmetic to structural reasoning. They avoid many bounds checks, but specifications often require recursive predicates for properties such as membership, length, occurrences, or
substructure. Proofs may then require helper contracts or lemmas that connect these recursive definitions to the implementation.

Control structure also matters. Recursive implementations often align with recursive specifications, allowing proof obligations to follow recursive calls. Imperative implementations must instead summarize mutable state with loop invariants: an accumulator, index, or reference cell may have an obvious algorithmic role, but that role must be made explicit for the verifier.

\begin{table}[t]
\centering
\caption{Typical proof obligations induced by implementation structure.}
\vspace{-2mm}
\label{tab:proof-burdens}
\begin{tabular}{p{0.27\linewidth} p{0.65\linewidth}}
\hline
\textbf{Program choice} & \textbf{Typical proof burden and guidance} \\
\hline
\textbf{Array + imperative}
& Bounds checks, index arithmetic, and quantified range facts; usually requires loop invariants over processed prefixes, suffixes, or windows, sometimes with ghost code to record processed-state summaries. \\

\textbf{Array + recursive}
& Bounds and index obligations remain, but recursive helper contracts can expose smaller subproblems and align proof obligations with recursive calls. \\

\textbf{List + recursive}
& Structural reasoning over \texttt{Nil}/\texttt{Cons}, length, membership, occurrences, and substructure; usually requires recursive predicates and helper lemmas. \\

\textbf{List + imperative}
& Mutable traversal plus structural list reasoning; usually requires invariants and ghost code relating visited elements, remaining elements, and logical summaries used in the specification. \\
\hline
\end{tabular}
\end{table}

Table~\ref{tab:proof-burdens} summarizes the typical proof obligations induced by the four implementation families studied in this paper.
These differences motivate implementation diversity. Two implementations may be intended to satisfy the same task-level semantics while producing substantially different verification conditions. A failed verification attempt may
therefore reflect an incorrect implementation, but it may also reflect a weak invariant, a missing termination argument, or a helper fact that has not been exposed to the solver. \toolname tests this hypothesis by generating
array-recursive, array-imperative, list-recursive, and list-imperative variants
for each benchmark task.

\section{Diversify2Verify by Example}
\label{sec:pipeline-overview}

\begin{figure}[t]
\centering
\resizebox{1\linewidth}{!}{%
\begin{tikzpicture}[
  node distance=0.65cm and 0.55cm,
  input/.style={
    draw,
    rounded corners,
    align=center,
    text width=2.25cm,
    minimum height=1.05cm,
    inner sep=5pt,
    font=\small,
    fill=inputgray
  },
  stage/.style={
    draw,
    rounded corners,
    align=center,
    text width=2.35cm,
    minimum height=1.05cm,
    inner sep=5pt,
    font=\small,
    fill=stageblue
  },
  check/.style={
    draw,
    rounded corners,
    align=center,
    text width=2.45cm,
    minimum height=1.05cm,
    inner sep=5pt,
    font=\small,
    fill=checkpurple
  },
  test/.style={
    draw,
    rounded corners,
    align=center,
    text width=2.15cm,
    minimum height=1.05cm,
    inner sep=5pt,
    font=\small,
    fill=testorange
  },
  success/.style={
    draw,
    rounded corners,
    align=center,
    text width=2.35cm,
    minimum height=0.9cm,
    inner sep=5pt,
    font=\small,
    fill=successgreen
  },
  failure/.style={
    draw,
    rounded corners,
    align=center,
    text width=2.35cm,
    minimum height=0.9cm,
    inner sep=5pt,
    font=\small,
    fill=failred
  },
  arrow/.style={
    -{Latex[length=2mm]},
    thick
  },
  repair/.style={
    -{Latex[length=2mm]},
    thick,
    dashed,
    draw=black!55
  }
]

\node[input] (input)
  {Task +\\Test Cases};

\node[stage, right=of input] (stageA)
  {\stageA (\S\ref{sec:contract-generation})\\Contract\\Generation};

\node[check, right=of stageA] (whyA)
  {Run \whythree\\on Spec. +\\Test Lemmas};

\node[stage, right=of whyA] (stageB)
  {\stageB (\S\ref{sec:implementation-generation})\\Implementation\\Generation};

\node[check, right=of stageB] (testsB)
  {Run \whythree\\Test Cases};

\node[stage, below=1.55cm of whyA] (stageC)
  {\stageC (\S\ref{sec:full-verification})\\Full\\Verification};

\node[check, right=of stageC] (whyC)
  {Run \whythree\\Full Verifier};

\node[success, above right=0.12cm and 0.65cm of whyC] (success)
  {$\checkmark$ Verified\\\whythree Artifact};

\node[failure, below right=0.12cm and 0.65cm of whyC] (failure)
  {$\times$ Unverified\\\whythree Artifact};

\node[font=\footnotesize, align=center, below=0.04cm of failure]
  {repair budget exhausted};

\draw[arrow] (input) -- (stageA);
\draw[arrow] (stageA) -- (whyA);
\draw[arrow] (whyA) -- (stageB);
\draw[arrow] (stageB) -- (testsB);

\draw[arrow]
  (testsB.south) -- ++(0,-0.35) -| (stageC.north);

\draw[arrow] (stageC) -- (whyC);

\draw[arrow]
  (whyC.east) -- ++(0.35,0) |- (success.west);

\draw[arrow]
  (whyC.east) -- ++(0.35,0) |- (failure.west);

\draw[repair]
  (whyA.north) to[out=110,in=70,looseness=1.15]
  node[above, font=\footnotesize, align=center]
  {repair spec.\\or test lemmas}
  (stageA.north);

\draw[repair]
  (testsB.north) to[out=110,in=70,looseness=1.15]
  node[above, font=\footnotesize, align=center]
  {repair or\\regenerate code}
  (stageB.north);

\draw[repair]
  (whyC.south) to[out=-110,in=-70,looseness=1.2]
  node[below, font=\footnotesize, align=center]
  {repair invariants,\\assertions, or lemmas}
  (stageC.south);

\end{tikzpicture}%
}
\vspace{-3mm}
\caption{Overview of the \toolname staged verification pipeline. \whythree is
used after \stageA to check the generated specification and test lemmas, during
\stageB to execute test cases against candidate implementations, and
after \stageC to discharge the verification conditions for the final annotated
artifact. 
}
\label{fig:pipeline-overview}
\end{figure}
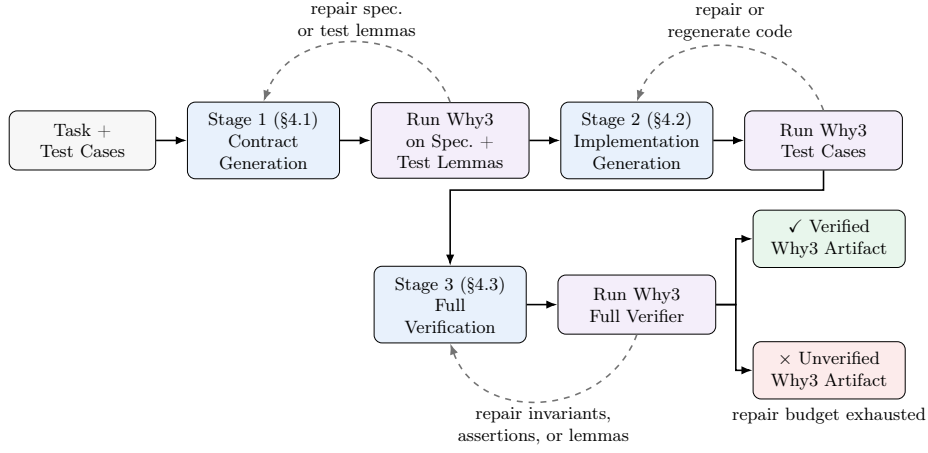

Figure~\ref{fig:pipeline-overview} shows the \toolname pipeline. Given a task description, target signature, concrete tests, and target representation, \toolname separates generation into three stages: \stageA infers a semantic contract, \stageB generates executable implementations, and \stageC adds proof annotations to produce a verification artifact.
This separation makes explicit the roles of the three artifacts. The contract states what makes a result correct, the implementation computes a candidate result, and the final artifact must provide the proof bridges that connect local program state to the contract.

\paragraph{Running example.}
We use \emph{Maximum Distance Between a Pair of Values} as a running example.
The function takes two non-increasing, zero-indexed integer arrays,
\texttt{nums1} and \texttt{nums2}. A pair of indices $(i,j)$ is valid when both
indices are in bounds, $i \leq j$, and \texttt{nums1[i] <= nums2[j]}. The
distance of a valid pair is $j-i$. The task is to return the maximum valid
distance, or $0$ when no valid pair has positive distance.
For example, given
\[
  \texttt{nums1 = [30,29,19,5]}
  \qquad
  \texttt{nums2 = [25,25,25,25,25]},
\]
the index pair $(i,j)=(2,4)$ is valid because
\texttt{nums1[2] = 19}, \texttt{nums2[4] = 25}, $2 \leq 4$, and
$19 \leq 25$. Its distance is $j-i = 4-2 = 2$. No valid pair has larger
distance, so the expected output is $2$.

\paragraph{Contract generation.}
\stageA defines correctness without describing how to compute the result. For the array version of the running example, the generated contract captures well-formed inputs, non-increasing arrays, valid index pairs, pair distances, and maximality. 

The key predicate is \texttt{maximum\_distance\_spec}. It separates the zero case from the positive case: a positive result must be realized by some valid pair and must upper-bound the distance of every valid pair. Thus, the contract
specifies the semantic property of the result, not the algorithm used to compute it. The full \stageA artifact also includes helper definitions and test lemmas derived from the concrete examples. These test lemmas ensure the contract aligns with the concrete examples, though passing them does not guarantee the natural-language task is perfectly captured.

\vspace{-1mm}
\begin{lstlisting}[
style=why3style,
label={lst:stage-a-contract}
]
predicate valid_pair
  (nums1: array int) (nums2: array int) (i: int) (j: int) =
  0 <= i < length nums1 /\ 0 <= j < length nums2 /\
  i <= j /\ nums1[i] <= nums2[j]

predicate has_pair_distance
  (nums1: array int) (nums2: array int) (d: int) =
  exists i j: int.
    valid_pair nums1 nums2 i j /\ d = pair_distance i j

predicate distance_upper_bound
  (nums1: array int) (nums2: array int) (d: int) =
  forall i j: int. valid_pair nums1 nums2 i j -> pair_distance i j <= d

predicate maximum_distance_spec
  (nums1: array int) (nums2: array int) (res: int) =
  (res = 0 /\ no_positive_distance_pair nums1 nums2) \/
  (0 < res /\ has_pair_distance nums1 nums2 res /\
   distance_upper_bound nums1 nums2 res)

val maximum_distance_between_a_pair_of_values_array
  (nums1: array int) (nums2: array int) : int
  requires { valid_input nums1 nums2 }
  ensures  { maximum_distance_spec nums1 nums2 result }
\end{lstlisting}
\vspace{-2mm}

\paragraph{Implementation generation.}
\stageB generates executable WhyML candidates for the same semantic task. For the running example, the array-imperative variant uses the standard two-pointer scan, maintaining indices \texttt{i} and \texttt{j} and a reference \texttt{best} for the largest distance found so far. When \texttt{nums1[i] <= nums2[j]}, the implementation updates \texttt{best} and advances \texttt{j}; otherwise, it advances \texttt{i}.

\paragraph{Full verification.}
\stageC combines an accepted contract with a tested implementation and adds the
annotations needed for \whythree verification. For the array-imperative variant,
the main challenge is to prove that the local two-pointer scan establishes the
global maximality condition required by
\texttt{maximum\_distance\_spec}. 

\begin{lstlisting}[
style=why3style,
label={lst:stage-c-invariants}
]
while !i < length nums1 && !j < length nums2 do
  invariant { 0 <= !i <= length nums1 }
  invariant { 0 <= !j <= length nums2 }
  invariant { 0 <= !best }
  invariant { !best = 0 \/ has_pair_distance nums1 nums2 !best }
  invariant { processed_j_upper_bound nums1 nums2 !j !best }
  invariant { left_future_excluded nums1 nums2 !i !j }
  variant   { (length nums1 - !i) + (length nums2 - !j) }
(* loop body *)
done
\end{lstlisting}

The initial invariants establish basic safety facts about the loop indices and the current value of \texttt{best}. The witness invariant connects \texttt{best} to the contract by requiring that any positive value of \texttt{best} is realized by some valid pair. The final two invariants provide
the main proof bridge. \texttt{processed\_j\_upper\_bound} states that all valid pairs with already processed right indices have distance at most \texttt{best}, while \texttt{left\_future\_excluded} justifies discarding left
indices that cannot produce better future pairs. Together, these annotations connect the local state of the two-pointer scan to the global maximality condition in the postcondition.

For this task, the array-imperative and array-recursive variants verify without repair, whereas the list-recursive and list-imperative variants require two repair attempts. Since the task is naturally array-oriented, this example should not be read as evidence that arrays are generally easier to verify. It instead illustrates the broader point studied in our evaluation: representation and control-structure choices can substantially change verification effort.

\section{Generation and Repair Policies}
\label{sec:method}

Section~\ref{sec:pipeline-overview} described the pipeline through one task. We now summarize the policies used across all benchmarks. These policies define what each stage may generate, what each stage may repair, and which semantic
boundaries must be preserved.

Table~\ref{tab:stage-policies} summarizes the policy enforced at each stage. The central rule is a frozen-contract discipline: once a \stageA contract is accepted for a representation, later stages may repair implementations and proof annotations, but they may not change the public semantic target for that representation.

\begin{table}[t]
\centering
\small
\caption{Repair boundaries in Diversify2Verify.}
\vspace{-2mm}
\label{tab:stage-policies}
\begin{tabular}{p{0.22\linewidth} p{0.72\linewidth}}
\hline
\textbf{Stage} & \textbf{Policy} \\
\hline
\textbf{\stageA:}\newline\textbf{Contract}
& Repair syntax/type errors, helper definitions, example lemmas, and semantic mistakes before freezing; the accepted representation-specific contract is then fixed. \\

\textbf{\stageB:}\newline\textbf{Implementation}
& Repair or regenerate executable code only within the same representation and control-structure family. \\

\textbf{\stageC:}\newline\textbf{Verification}
& Add proof scaffolding such as invariants, assertions, ghost code, internal helper contracts, and lemmas; preserve the contract, implementation family, and executable behavior.
\\
\hline
\end{tabular}
\vspace{-5mm}
\end{table}
\subsection{Contract generation policy}
\label{sec:contract-generation}

\stageA takes a natural-language problem statement, a target signature, concrete examples, and a target representation. It produces a representation-specific Why3 contract containing the target declaration, preconditions, postconditions, and any helper predicates or logical functions needed to express the task.

The natural-language description is treated as the primary source of semantic intent. Examples are used as concrete checks and as limited evidence for resolving ambiguity, but not as a complete behavioral specification. \stageA
therefore separates assumptions about valid inputs from requirements on the output. Input assumptions become \texttt{requires} clauses or explicit validity predicates, while output requirements become \texttt{ensures} clauses.

The contract is written in the reasoning style of the target representation. Array contracts use lengths, explicit bounds, and bounded quantification over indices. List contracts use structural definitions over \texttt{Nil} and
\texttt{Cons}, or standard list-library notions when they match the task. The goal is not to mechanically translate between arrays and lists, but to produce representation-appropriate contracts that capture the same underlying task. Thus, we use task-equivalent to mean that the array and list variants target the same benchmark-level semantics through representation-specific contracts; we do not additionally prove that the two formal contracts are logically equivalent.

\stageA favors direct mathematical characterizations over executable-looking specifications. For optimization tasks, the postcondition usually separates feasibility from optimality: the result must be achieved by a valid witness, and no valid witness may have a better value. For Boolean, counting, and transformation tasks, the postcondition states the exact semantic relation between the inputs and the returned value.

After generation, the \stageA artifact is first parsed and type-checked by Why3. We then run a separate \stageA validation pass that attempts to prove the example-derived test lemmas. These lemmas instantiate the generated contract on concrete benchmark examples and serve as a lightweight consistency check between the contract and the examples. They do not prove that the contract fully captures the natural-language task. We therefore use accepted contract to mean a representation-specific contract that is well formed in Why3 and whose example-derived test lemmas are discharged by the \stageA validation procedure, possibly after one repair attempt. Acceptance is not a proof of full semantic correctness with respect to the natural-language task.

Repair is guided by the kind of \stageA failure. Syntax and type errors are fixed directly. If the artifact type-checks but the separate \stageA validation pass fails to prove the example-derived lemmas, its diagnostics are passed to at most one \stageA repair call. This repair may add assertions, unfold helper definitions, or simplify logically equivalent helper predicates. Semantic repair is allowed only before freezing, and only when the contract is inconsistent with the task statement or examples. Once the artifact is well formed and its example-derived lemmas are proved, the accepted contract is frozen for later stages.

\subsection{Implementation diversity policy}
\label{sec:implementation-generation}

\stageB generates executable WhyML candidates. 
For each benchmark task, \toolname generates four implementation families: array-recursive, array-imperative, list-recursive, and list-imperative. Recursive variants use structural recursion or recursive helpers. Imperative variants use loops, mutable references, and explicit state updates.

Within a fixed family, the implementation may choose the algorithmic strategy, traversal order, helper decomposition, and accumulator structure. During repair, however, it must preserve the accepted \stageA contract, the requested
representation, and the requested control-structure family. Thus, an array-imperative candidate may be repaired into a different array-imperative algorithm, but it may not become list-based or recursive, and it may not modify the contract it is checked against.

Diversity is introduced to preserve the proof-obligation differences summarized in Table~\ref{tab:proof-burdens}. By generating one recursive and one imperative implementation for each representation, the evaluation can measure how data representation and control structure affect verifiability while keeping the accepted \stageA contract fixed within each representation.

Each candidate is checked by type checking and by executable tests derived from the benchmark examples. Failed candidates are repaired or regenerated within a bounded budget of five attempts while preserving their representation and implementation style. Passing these tests does not establish correctness; it only determines which candidates are submitted to \stageC.

\subsection{Final verification repair policy}
\label{sec:full-verification}

\stageC turns one tested implementation into a deductive verification artifact. It combines the \stageA contract with the \stageB implementation and removes the executable tests, which are no longer part of the final proof artifact.
It then adds proof scaffolding such as internal helper contracts, loop invariants, variants, assertions, ghost code, and auxiliary lemmas. These annotations may strengthen the proof context, but they may not weaken or
replace the \stageA contract.

The main proof obligation is to bridge local program state to the semantic contract generated in \stageA. Local state includes loop indices, recursive arguments, accumulators, partial results, and mutable references. \stageC annotations are therefore proof guidance: they expose the relationship between these local quantities and the specification without changing the meaning of the program.

The generated verification conditions cover safety, termination, and functional correctness. Safety obligations prove that operations such as array accesses and
pattern matches are well defined. Termination obligations are discharged using loop or recursion variants. Functional-correctness obligations require the proof scaffolding to imply the \stageA postconditions at loop exits, recursive base cases, and helper returns.

If verification fails, \stageC repair is restricted to proof-oriented changes using two repair attempts. Typical repairs strengthen loop invariants, add missing variants, insert local assertions, introduce ghost code, or prove helper lemmas that expose facts needed by the SMT solvers. Local implementation edits are permitted only when they are semantics-preserving refactorings or proof-enabling local changes that preserve the implementation family and executable behavior on valid inputs. \stageC may not weaken the \stageA contract, add stronger preconditions to the target entry point, change the target signature, switch between array and list representations, or replace a recursive implementation with an imperative one or vice versa to make verification easier.

These restrictions are imposed by the \stageC prompts and repair driver, but the LLM is not a trusted enforcement mechanism. We therefore treat preservation of the frozen \stageA contract as a property to be checked after generation, rather than as something guaranteed by the prompt alone.
A successful \stageC run produces a fully verified Why3 artifact. A failed run produces the best unverified artifact together with verifier diagnostics and
repair traces.

\section{Benchmark Construction}
\label{sec:benchmark}

We construct the benchmark from LeetCodeDataset~\cite{xia2025leetcodedataset}, which provides Leet\-Code-style Python programming problems with natural-language descriptions and executable test cases. Because LeetCode problems are public, we do not claim that this benchmark is contamination-free or that it measures a model's ability to solve unseen programming tasks. The descriptions, standard algorithms, or Python solutions may have appeared in model training data. We instead use LeetCodeDataset as a source of task descriptions and tests for constructing a controlled verification benchmark. The evaluated artifacts---Why3 contracts, WhyML implementations, proof annotations, repair traces, and verifier outcomes---are generated by our pipeline and are not part of the original dataset. While potential contamination might help the model recognize the underlying algorithmic task, our main measurement is comparative: whether different representation and control-structure choices lead to different verification outcomes under the same verifier and repair budget.

\begin{table}[t]
\centering
\caption{Summary of the verification benchmark.}
\vspace{-2mm}
\label{tab:benchmark-summary}
\begin{tabular}{lr}
\toprule
Property & Value \\
\midrule
Unique tasks & 73 \\
Available tests per task & min 25, median 109, mean 106.7, max 178 \\
Input arity & 30 one-arg, 34 two-arg, 7 three-arg, 2 four-arg \\
Return shape & 58 integer, 15 array or list \\
\bottomrule
\end{tabular}
\vspace{-3mm}
\end{table}
Starting from roughly 2,900 tasks, we first filtered for problems over primitive values and integer sequences, such as arrays and lists, excluding strings, trees, floating-point values, and other structures that would require additional Why3 modeling. This type-based pruning produced approximately 1,000 candidates. We then used an LLM-assisted triage pass to retain tasks whose behavior appeared expressible with arithmetic and integer-sequence properties and that admitted multiple plausible implementation strategies, reducing the set to around 150 benchmarks.
Finally, we applied the \stageA contract-generation and validation procedure from Section~\ref{sec:contract-generation}. We retained only benchmarks for which \stageA produces accepted contracts for the target representations used in the evaluation. As in Section~\ref{sec:contract-generation}, acceptance means that the contract is well formed in Why3 and that the example-derived validation lemmas are discharged, possibly after repair. This filter yielded 73 benchmark tasks and is part of the benchmark construction process rather than an independent \stageA success result.

Table~\ref{tab:benchmark-summary} summarizes the final benchmark. The retained set contains 73 tasks with varying input arities, return shapes, and test-suite sizes. For each task, \toolname generates four implementation variants, yielding 292 artifacts. The benchmark should therefore be read as a controlled verification benchmark, not as a representative sample of arbitrary LeetCode tasks: it is intentionally biased toward SMT-friendly data types, tractable specifications, and tasks with accepted \stageA contracts. This selection lets us isolate the paper's main question: how representation and control-structure choices affect deductive verifiability.

\section{Implementation}
\label{sec:implementation}

\toolname is implemented as a scripted, non-interactive pipeline around Codex. We use Codex as the programming agent: it receives a stage prompt, applies the local stage instructions, creates or edits WhyML files, and returns a candidate artifact. The underlying model configuration is selected per stage. Stages~1 and~3 use GPT-5.5 with high reasoning, because these stages involve formal specification and proof construction. \stageB uses GPT-5.4 with medium
reasoning, because this stage primarily generates executable implementations that are later checked against the accepted contract.

Each stage is invoked by a driver script. The driver builds a prompt from the benchmark description, target signature, examples, target representation, and, when applicable, the implementation family. It then invokes Codex, extracts the
WhyML artifact, runs the checks, and either accepts the artifact or launches a bounded repair attempt. The drivers are non-interactive: all inputs, model configurations, checker commands, and repair budgets are fixed
before evaluation.

The Codex context includes a local skill for each stage. A skill is a stage-specific instruction file that describes the expected WhyML artifact shape, permitted Why3 libraries, representation-specific idioms, checker commands, and common failure patterns. The \stageA skills focus on contract structure, helper predicates, and example-derived test lemmas. The \stageB skills focus on executable recursive or imperative WhyML code for the selected representation. The \stageC skills focus on proof-oriented annotations, including invariants, variants, assertions, ghost state, internal helper contracts, and auxiliary lemmas.

During generation, each driver performs only the checks needed to reject malformed artifacts early. \stageA contracts and \stageC verification artifacts must parse and type-check in Why3. \stageB implementations must additionally pass the selected executable tests. These fast checks prevent invalid artifacts from flowing to later stages, but they do not require all proof obligations to be discharged. 

\begin{table}[t]
\centering
\small
\setlength{\tabcolsep}{3pt}
\renewcommand{\arraystretch}{1.12}
\caption{Generation configuration used by \toolname. }
\vspace{-2mm}
\label{tab:implementation-config}
\begin{tabularx}{\linewidth}{
  @{}
  >{\raggedright\arraybackslash}p{0.21\linewidth}
  >{\raggedright\arraybackslash}X
  >{\raggedright\arraybackslash}p{0.19\linewidth}
  >{\raggedright\arraybackslash}p{0.13\linewidth}
  @{}
}
\hline
\textbf{Stage} & \textbf{Task and feedback} & \textbf{Model} & \textbf{Budget} \\
\hline
\makecell[tl]{\textbf{\stageA:}\\\textbf{Contract}} &
Generate a representation-specific contract; type-check it and validate example lemmas. &
GPT-5.5 (high) &
1 repair \\

\makecell[tl]{\textbf{\stageB:}\\\textbf{Implementation}} &
Generate executable WhyML; check with type checking and executable tests. &
GPT-5.4 (medium) &
5 attempts \\

\makecell[tl]{\textbf{\stageC:}\\\textbf{Verification}} &
Add proof annotations; repair using Why3 verification diagnostics. &
GPT-5.5 (high) &
2 repairs \\
\hline
\end{tabularx}
\vspace{-3mm}
\end{table}

For each representation-specific \stageA contract, the validation pass attempts to prove seven example-derived lemmas. For \stageB, each candidate implementation is checked on 20 executable tests selected from the benchmark test suite. These tests are used only as filtering checks before \stageC; they are not part of the final deductive verification artifact.

Full verification is run by a separate Why3 verifier driver rather than by Codex itself. In our experiments, we used Why3 1.8.2 together with the SMT solvers Z3 4.8.6, Alt-Ergo 2.4.0, and CVC4 1.8, using a 10-second timeout per solver call. The verifier first attempts direct proving with this solver portfolio. For goals that remain, it applies increasingly aggressive Why3 transformations, first splitting large verification conditions and then inlining helper definitions when needed. An artifact is counted as verified only if all generated verification conditions are discharged by this verifier configuration. When \stageC verification fails, the verifier diagnostics are passed back to Codex only as feedback for the bounded repair calls summarized in Table~\ref{tab:implementation-config}.

When verification or filtering fails, \toolname relies on bounded, log-guided Codex calls for repair (summarized in Table~\ref{tab:implementation-config}). Each repair prompt includes the failed artifact, the relevant checker or verifier diagnostics, and the stage inputs needed to preserve context. For \stageC, this includes the failing \whythree file, the verifier log, the benchmark description, the frozen \stageA specification, and the tested \stageB implementation.
The first repair attempt addresses the initial failure. Later attempts use the updated diagnostics to focus on the remaining unproved goals. Each repaired artifact is then rechecked by the corresponding driver.

\section{Evaluation}
\label{sec:evaluation}

Our evaluation focuses on \stageC, where \toolname attempts to turn executable WhyML implementations into deductively verified artifacts. \stageA and \stageB serve as prerequisite filters for this experiment. \stageA admits tasks for which both array and list contracts are accepted, and \stageB admits only implementations that pass type checking and executable tests. The main evaluation therefore begins after these filters: \emph{given an accepted contract and a diverse set of tested implementations, can \stageC produce fully verified artifacts, and which representation or implementation style is easiest to verify}?

\paragraph{Pipeline setup results.}
The benchmark contains 73 tasks after the \stageA acceptance filter described in Section~\ref{sec:benchmark}. Because each retained task has accepted array and list contracts by construction, the evaluation uses 146 representation-specific \stageA contracts. Of these accepted contracts, 85 required no repair and 61 required one contract-repair attempt. The array contracts account for 45 initially accepted contracts and 28 accepted after repair, while the list contracts account for 40 initially accepted contracts and 33 accepted after repair.

\stageB targets four variants per task: array-recursive, array-imperative, list-recursive, and list-imperative. Within the five-attempt \stageB budget, \toolname obtains one type-checked, test-passing implementation for each variant, yielding 292 candidates for \stageC. The main evaluation therefore centers on the 292 final verification artifacts produced by combining these tested implementations with the accepted \stageA contracts.

\paragraph{Research Questions.} We evaluate three research questions:

\begin{description}
\item[RQ1: Verification effectiveness.]
How often does \stageC verify  the implementations, and how much do repair passes improve the verification rate?

\item[RQ2: Representation and implementation diversity.]
How does verification success vary across array and list representations and across recursive and imperative implementations? Does generating multiple variants increase the chance of obtaining at least one verified artifact for a
task?

\item[RQ3: Remaining proof failures.]
For artifacts that remain unverified after repair, what kinds of proof obligations are left?
\end{description}

\subsection{RQ1: Verification effectiveness}
\label{sec:evaluation-rq1}

Table~\ref{tab:stage3-overall} summarizes the overall \stageC results. Of the 292 verification artifacts, 96 verify as generated. The first Codex-assisted repair pass verifies 35 additional artifacts, and the second repair pass
verifies 23 more. Overall, \toolname verifies 154 artifacts after repair, corresponding to a final artifact-level verification rate of 52.7\%.

\begin{table}[t]
\centering
\caption{Overall \stageC verification results. Repair increases the number of
verified artifacts from 96 initially to 154 after two repair passes.}
\vspace{-2mm}
\begin{tabular}{lrrrr}
\toprule
Attempt & Newly verified & Cumulative verified & Remaining & Cumulative rate \\
\midrule
Initial \stageC & 96 & 96 & 196 & 32.9\% \\
After repair 1 & 35 & 131 & 161 & 44.9\% \\
After repair 2 & 23 & 154 & 138 & 52.7\% \\
\bottomrule
\end{tabular}
\label{tab:stage3-overall}
\vspace{-2mm}
\end{table}

Repair is therefore responsible for a substantial fraction of the final successes. Among the 154 verified artifacts, 58 require at least one repair pass. The first repair pass gives the larger improvement, but the second pass still verifies 23 artifacts that remain unproved after the first repair.

Most artifacts are proved without requiring the most aggressive \whythree transformations. Among the 154 verified artifacts, 133 are proved directly, 13 require splitting verification conditions, and 8 require both splitting and inlining helper definitions. This suggests that once the right contracts, invariants, and assertions are present, most verified artifacts are within reach of the base SMT portfolio.

\subsection{RQ2: Representation and implementation diversity}
\label{sec:evaluation-rq2}

At the task level, diversity increases the chance of obtaining a verified solution. For each task, we generate four variants, corresponding to the cross-product of representation and implementation style: array-recursive,
array-imperative, list-recursive, and list-imperative.
Table~\ref{tab:task-level-diversity} groups the 73 tasks by how many of these variants verify after repair. At least one variant verifies for 49 tasks, giving a task-level success rate of 67.1\%. Among these 49 solved tasks, 25 are solved by only a subset of the variants, showing that different implementations make different tasks amenable to verification.

Shifting from task-level success to individual artifact performance, Table~\ref{tab:stage3-by-variant} breaks down verification by representation and implementation style. Overall, recursive implementations verify more often than imperative implementations. Across both representations, 84 of 146 recursive artifacts verify, compared with 70 of 146 imperative artifacts. Array and list representations are closer overall: 79 of 146 array artifacts verify, compared
with 75 of 146 list artifacts. The strongest individual variant family is array-recursive with 44 verified artifacts, followed closely by list-recursive with 40. Both imperative families trail behind, verifying 35 artifacts each.

\begin{table}[t]
\centering
\caption{Task-level verification outcomes over the 73 benchmark tasks. A task
is counted as solved if at least one of its four variants verifies after
repair.}
\begin{tabular}{lrr}
\toprule
Task-level outcome & Tasks & \% of tasks \\
\midrule
No variant verified & 24 & 32.9\% \\
Exactly one variant verified & 5 & 6.8\% \\
Exactly two variants verified & 7 & 9.6\% \\
Exactly three variants verified & 13 & 17.8\% \\
All four variants verified & 24 & 32.9\% \\
\midrule
At least one variant verified & 49 & 67.1\% \\
\bottomrule
\end{tabular}
\label{tab:task-level-diversity}
\end{table}

\begin{table}[t]
\centering
\caption{\stageC verification results by representation and implementation style. Recursive variants verify more often than imperative variants, while array and list representations have similar final verification rates.}
\begin{tabular}{lrrrrrrr}
\toprule
Variant & Total & Initial & Repair 1 & Repair 2 & Verified & Unverified & Rate \\
\midrule
Array recursive & 73 & 28 & 11 & 5 & 44 & 29 & 60.3\% \\
Array imperative & 73 & 21 & 7 & 7 & 35 & 38 & 47.9\% \\
List recursive & 73 & 30 & 7 & 3 & 40 & 33 & 54.8\% \\
List imperative & 73 & 17 & 10 & 8 & 35 & 38 & 47.9\% \\
\midrule
Total & 292 & 96 & 35 & 23 & 154 & 138 & 52.7\% \\
\bottomrule
\end{tabular}
\label{tab:stage3-by-variant}
\vspace{-3mm}
\end{table}

This difference is not surprising in a deductive-verification setting. Many \stageA specifications are expressed using recursive logical functions or inductive predicates. Recursive implementations often follow the same structure, so the proof can proceed by exposing the contract of each recursive call. In contrast, imperative implementations require loop invariants that summarize the
entire processed portion of the input and relate mutable program state to the logical specification. These invariants are often the hardest part of the proof
to synthesize automatically.

The repair results suggest that imperative implementations are not worse executable candidates; they are harder to verify automatically. Before repair, recursive variants verify substantially more often than imperative variants. Repair narrows this gap by verifying 26 additional recursive artifacts and 32 additional imperative artifacts. This suggests that many imperative failures are due to missing proof structure, such as loop invariants, variants, and bridge assertions, rather than an incorrect implementation strategy.

Thus, diversity helps not because one representation dominates, but because different variants expose different proof opportunities. Recursive implementations often align better with recursive specifications, while imperative ones can become competitive once repair adds the missing loop-level proof structure.

\subsection{RQ3: Remaining proof failures}
\label{sec:evaluation-rq3}
After two repair passes, 138 artifacts remain unverified, even though all corresponding \stageB implementations passed type checking and executable tests. These failures are therefore not failures to obtain type-checked, test-passing candidates; they arise at the deductive verification stage.

Many of these failures are close to verification. Among the 138 unverified artifacts, 62 have only one remaining unproved root goal, and 110 have at most three. The average number of remaining root goals is 2.5. This suggests that many failures are localized proof gaps rather than evidence of fundamentally incorrect implementations.

The array-imperative artifact for the \emph{Longest Turbulent Subarray} benchmark illustrates this pattern. The task is to return the length of the longest contiguous segment whose adjacent comparisons strictly alternate; equal adjacent elements break the segment. The implementation performs the standard linear scan, tracking the current turbulent suffix and the best length seen so far. Its last remaining obligation arises when the current comparison alternates with the previous one, so the suffix is extended. The missing bridge is that any turbulent interval ending at \texttt{i+1} can be shortened to one ending at \texttt{i}; the previous maximality invariant then bounds its length by \texttt{old\_current + 1}. We confirmed this diagnosis in a manual follow-up: adding proof-only helper lemmas for suffix-shortening and for lifting the previous bound to all intervals ending at \texttt{i+1} discharged the remaining goal, without changing the executable code or weakening the specification. This repair was not included in the reported results, but it shows that at least some failures are due to missing proof structure rather than incorrect artifacts.

This example illustrates a common remaining failure mode: the generated proof scaffolding does not expose the semantic bridge needed by the SMT solvers. Across the remaining failures, we see similar gaps involving insufficient loop invariants, missing helper contracts, array bounds or index arithmetic, structural list reasoning, and reasoning about counting, cardinality, or optimality.

Overall, the remaining failures suggest two directions for improvement. Repair prompts could target bridge obligations more directly, such as connecting helper postconditions, loop invariants, or accumulator meanings to the specification. The pipeline could also include specialized proof patterns for recurring domains such as index-range reasoning, counting, and optimization.

\subsection{Discussion and limitations}
\label{sec:evaluation-discussion}

The \stageC results show that both repair and implementation diversity matter. Repair increases artifact-level verification from 96 of 292 artifacts, or 32.9\%, to 154 of 292 artifacts, or 52.7\%. At the task level, using all four variants yields at least one verified artifact for 49 of 73 tasks, or 67.1\%, compared with 44 tasks for the strongest individual family.

Importantly, these results are relative to the accepted \stageA contracts. Deductive verification proves that a \stageC implementation satisfies its contract, not that the contract fully captures the natural-language benchmark description. To assess this specification risk, we ran a screening audit using GPT-5.5 (configured for high reasoning) on the 292 final artifacts. The judge evaluated two criteria: whether the \stageA specification matched the benchmark description, and whether \stageC preserved that specification.

\paragraph{Specification Match} The judge flagged 10 potential mismatches. Because the 146 \stageA contracts are reused across variants, we manually inspected these flagged cases to find the root cause. Only one case corresponded to a specification that was genuinely too weak and could potentially accept incorrect programs. Seven cases involved missing input-bound constraints typical of LeetCode problems; however, these omissions did not affect the intended behavior on valid inputs. The remaining two cases involved inconsistencies between generated tests and stated input constraints; these tests were preemptively excluded before \stageB testing.

\paragraph{Specification Preservation} The judge flagged only one artifact for failing to preserve the contract. Here, \stageC added preconditions that were implied by the benchmark description but absent from the frozen \stageA contract—a clear violation of our strict \stageC policy. Overall, this 291-of-292 success rate demonstrates that the pipeline reliably preserves the boundary between specification generation and proof repair.

\paragraph{Limitations} Several threats to validity remain. First, the LLM judge is an audit mechanism, not a formal guarantee; subtle specification errors may persist. Second, all generation, repair, and evaluation steps use GPT-family models; other LLMs may yield different verification and repair behaviors. Third, passing \stageB's finite executable tests does not establish functional correctness, which is only attempted in \stageC. Fourth, because the benchmark is filtered through \stageA, our results specifically measure verification on tasks where both array and list contracts were successfully accepted. Finally, our focus on integer and sequence tasks means these conclusions may not directly transfer to programs requiring richer heap structures, floating-point arithmetic, concurrency, or external libraries. Furthermore, some remaining failures likely reflect limitations of the SMT portfolio rather than fundamentally unprovable implementations.

\section{Related Work}
\label{sec:related}

\paragraph{LLM-based code generation.}
Large language models have been widely studied for generating executable programs from natural-language prompts. Benchmarks and systems such as Codex
and HumanEval~\cite{humaneval}, MBPP~\cite{mbpp}, and AlphaCode~\cite{alphacode} established test-based functional correctness as the dominant evaluation criterion for LLM
code generation. In this setting, diversity is usually a sampling mechanism: generating many candidates increases the chance that one passes the available tests. \toolname uses tests only as an intermediate filter. The final artifact must also include a formal contract and enough proof guidance for Why3 to prove that the implementation satisfies that contract. Thus, our use of diversity is different from pass@k-style sampling: we study whether alternative representations and control structures lead to verification conditions that are easier or harder for SMT-backed deductive verification.

\paragraph{Deductive verification and formal proof systems.}
\toolname is built on Why3~\cite{filliatre2013why3} and is related to other deductive verification systems such as Dafny~\cite{dafny}, Viper~\cite{viper}, and Verus~\cite{verus}, as well as proof assistants such as Lean~\cite{lean} and Rocq~\cite{rocq}. These systems provide
languages for programs, specifications, and proofs, but also expose the central challenge addressed in this paper: verified programming requires not only code, but also contracts, invariants, lemmas, ghost state, and other proof structure. We use Why3 because WhyML supports both functional and imperative programming while making the separation between executable code, specifications, and proof artifacts explicit. Although other deductive verifiers could serve this role, Why3 allows us to compare recursive and imperative implementations over arrays and lists within a single, unified framework. Our goal is not to introduce a new verifier, but to study how LLM-generated program variants interact with an existing deductive verifier.

\paragraph{LLMs for formal reasoning and verified program synthesis.}
Recent work has used LLMs for autoformalization, theorem proving, and verifier-guided program generation. Autoformalization translates informal mathematical statements into formal languages~\cite{autoformalization1,autoformalization2}, while systems such as LeanDojo~\cite{leandojo} and Baldur~\cite{baldur} use LLMs together with proof-assistant feedback, retrieval, or automated provers to construct formal proofs. Other systems integrate LLMs directly with program verifiers. A large body of work focuses on Dafny, including systems that generate assertions (Laurel~\cite{laurel}, Daisy~\cite{daisy}), evaluate annotation generation (DafnyBench~\cite{dafnybench}), study AI-assisted method synthesis (Misu et al.~\cite{dafny-ai}), and check consistency between code and docstrings (Clover~\cite{clover}). Similar efforts target Rust via Verus (AutoVerus~\cite{autoverus}) or integrate LLMs broadly with automated reasoners (Lemur~\cite{lemur}).\toolname is complementary to these systems. Rather than asking only whether an LLM can produce one verified program or missing annotations for a fixed program, we ask whether the form of the implementation affects verifiability. 

\paragraph{Specification generation, proof guidance, and repair.}
Our first stage builds on prior work translating informal intent into formal postconditions~\cite{postconditions} and LLM-based specification generation with verification-guided selection~\cite{specgen}. These systems highlight a central risk in LLM-assisted verification: a specification can be verifiable while failing to capture the intended behavior. \toolname mitigates this risk by using examples to check the specification against the natural-language task and by using an LLM judge, followed by manual inspection, to analyze cases where the specification may not match the intended semantics.

Our final stage is related to invariant generation and proof guidance. Classical tools such as Daikon~\cite{daikon} dynamically infer likely invariants, while recent work studies LLM generation and ranking of loop invariants~\cite{invariants2,invariants3,invariants4} and neural synthesis for SMT-assisted proof-oriented programming~\cite{smt-proofs}. \toolname also performs bounded repair, but with stage-specific boundaries: \stageA may repair the specification before freezing, \stageB may regenerate code only within the chosen representation and family, and \stageC may refine proof scaffolding while preserving the contract and implementation family. These boundaries prevent verification from succeeding by weakening the specification or replacing a difficult implementation with an easier one.

\section{Conclusion}
\label{sec:conclusion}
Generating verified programs from natural language requires more than inferring a correct specification: it also requires an implementation whose structure is amenable to deductive verification. This paper introduced \toolname, a staged LLM-based approach that exploits this observation by generating multiple task-equivalent implementations across representation and control-structure families and checking them against fixed representation-specific contracts.

The main result is that implementation diversity improves task coverage. Across 73 benchmark tasks and 292 generated artifacts, \toolname verifies 96 artifacts initially(32.9\%), and 154 after two repair passes(52.7\%). At the task level, at least one variant verifies for 49 tasks, yielding a 67.1\% success rate. Recursive variants verify more often than imperative variants, while array and
list representations have similar aggregate rates. These results demonstrate that task-equivalent implementations can differ substantially in verifiability, and that generating diverse variants increases the chance of finding one a verifier can prove correct.

Future work should extend this diversity-oriented methodology to richer data structures, strings, floating-point arithmetic, and library-based programs. It should also explore stronger invariant generation, better failure classification, and formal equivalence between representation-specific contracts. Applying this approach to other systems, such as Verus, would help distinguish Why3-specific engineering effects from a broader principle: implementation diversity is a powerful, practical tool for LLM-assisted verified programming.

\section*{Acknowledgments} 

This work was partially supported by the National Science
Foundation (NSF) under Award CCF2427581 and DARPA under Agreement FA8750-24-9-1000.

\bibliographystyle{splncs04}
\bibliography{bibtex}

@string{esop = {ESOP}}

@string{lpar = {International conference on logic for programming artificial intelligence and reasoning}}

@string{icse = {International Conference on Software Engineering}}

@string{cade = {International Conference on Automated Deduction}}

@string{iclr = {ICLR}}

@incollection{viper,
  author       = {Peter M{\"{u}}ller and
                  Malte Schwerhoff and
                  Alexander J. Summers},
  title        = {{Viper: {A} Verification Infrastructure for Permission-Based Reasoning}},
  booktitle    = {Dependable Software Systems Engineering},
  pages        = {104--125},
  publisher    = {{IOS} Press},
  year         = {2017},
}

@article{daikon,
  author       = {Michael D. Ernst and
                  Jeff H. Perkins and
                  Philip J. Guo and
                  Stephen McCamant and
                  Carlos Pacheco and
                  Matthew S. Tschantz and
                  Chen Xiao},
  title        = {{The Daikon system for dynamic detection of likely invariants}},
  journal      = {Sci. Comput. Program.},
  volume       = {69},
  number       = {1-3},
  pages        = {35--45},
  year         = {2007}, 
}

@article{humaneval,
  author       = {Mark Chen and
                  Jerry Tworek and
                  Heewoo Jun and
                  Qiming Yuan and
                  Henrique Pond{\'{e}} de Oliveira Pinto and
                  Jared Kaplan and
                  Harri Edwards and
                  Yuri Burda and
                  Nicholas Joseph and
                  Greg Brockman and
                  Alex Ray and
                  Raul Puri and
                  Gretchen Krueger and
                  Michael Petrov and
                  Heidy Khlaaf and
                  Girish Sastry and
                  Pamela Mishkin and
                  Brooke Chan and
                  Scott Gray and
                  Nick Ryder and
                  Mikhail Pavlov and
                  Alethea Power and
                  Lukasz Kaiser and
                  Mohammad Bavarian and
                  Clemens Winter and
                  Philippe Tillet and
                  Felipe Petroski Such and
                  Dave Cummings and
                  Matthias Plappert and
                  Fotios Chantzis and
                  Elizabeth Barnes and
                  Ariel Herbert{-}Voss and
                  William Hebgen Guss and
                  Alex Nichol and
                  Alex Paino and
                  Nikolas Tezak and
                  Jie Tang and
                  Igor Babuschkin and
                  Suchir Balaji and
                  Shantanu Jain and
                  William Saunders and
                  Christopher Hesse and
                  Andrew N. Carr and
                  Jan Leike and
                  Joshua Achiam and
                  Vedant Misra and
                  Evan Morikawa and
                  Alec Radford and
                  Matthew Knight and
                  Miles Brundage and
                  Mira Murati and
                  Katie Mayer and
                  Peter Welinder and
                  Bob McGrew and
                  Dario Amodei and
                  Sam McCandlish and
                  Ilya Sutskever and
                  Wojciech Zaremba},
  title        = {Evaluating Large Language Models Trained on Code},
  journal      = {CoRR},
  volume       = {abs/2107.03374},
  year         = {2021},
  eprinttype   = {arXiv},
  eprint       = {2107.03374},
}

@article{xia2025leetcodedataset,
  author       = {Yunhui Xia and
                  Wei Shen and
                  Yan Wang and
                  Jason Klein Liu and
                  Huifeng Sun and
                  Siyue Wu and
                  Jian Hu and
                  Xiaolong Xu},
  title        = {LeetCodeDataset: {A} Temporal Dataset for Robust Evaluation and Efficient
                  Training of Code LLMs},
  journal      = {CoRR},
  volume       = {abs/2504.14655},
  year         = {2025},
  eprinttype   = {arXiv},
}

@article{dafny-ai,
  author       = {Md Rakib Hossain Misu and
                  Cristina V. Lopes and
                  Iris Ma and
                  James Noble},
  title        = {Towards AI-Assisted Synthesis of Verified Dafny Methods},
  journal      = {Proc. {ACM} Softw. Eng.},
  volume       = {1},
  number       = {{FSE}},
  pages        = {812--835},
  year         = {2024},
}

@article{invariants2,
  author       = {Saikat Chakraborty and
                  Shuvendu K. Lahiri and
                  Sarah Fakhoury and
                  Madanlal Musuvathi and
                  Akash Lal and
                  Aseem Rastogi and
                  Aditya Senthilnathan and
                  Rahul Sharma and
                  Nikhil Swamy},
  title        = {{Ranking LLM-Generated Loop Invariants for Program Verification}},
  journal      = {CoRR},
  volume       = {abs/2310.09342},
  year         = {2023},
  eprinttype   = {arXiv},
}

@inproceedings{smt-proofs,
  author       = {Saikat Chakraborty and
                  Gabriel Ebner and
                  Siddharth Bhat and
                  Sarah Fakhoury and
                  Sakina Fatima and
                  Shuvendu K. Lahiri and
                  Nikhil Swamy},
  title        = {{Towards Neural Synthesis for SMT-Assisted Proof-Oriented Programming}},
  booktitle    = {ICSE},
  pages        = {1755--1767},
  publisher    = {{IEEE}},
  year         = {2025},
}

@article{verus,
  author       = {Andrea Lattuada and
                  Travis Hance and
                  Chanhee Cho and
                  Matthias Brun and
                  Isitha Subasinghe and
                  Yi Zhou and
                  Jon Howell and
                  Bryan Parno and
                  Chris Hawblitzel},
  title        = {Verus: Verifying Rust Programs using Linear Ghost Types},
  journal      = {Proc. {ACM} Program. Lang.},
  volume       = {7},
  number       = {{OOPSLA1}},
  pages        = {286--315},
  year         = {2023},
}

@article{dafny-annotator,
  author       = {Gabriel Poesia and
                  Chloe Loughridge and
                  Nada Amin},
  title        = {dafny-annotator: AI-Assisted Verification of Dafny Programs},
  journal      = {CoRR},
  volume       = {abs/2411.15143},
  year         = {2024},
  eprinttype   = {arXiv},
  eprint       = {2411.15143},
}

@article{invariants3,
  author       = {Adharsh Kamath and
                  Aditya Senthilnathan and
                  Saikat Chakraborty and
                  Pantazis Deligiannis and
                  Shuvendu K. Lahiri and
                  Akash Lal and
                  Aseem Rastogi and
                  Subhajit Roy and
                  Rahul Sharma},
  title        = {Finding Inductive Loop Invariants using Large Language Models},
  journal      = {CoRR},
  volume       = {abs/2311.07948},
  year         = {2023},
  eprinttype   = {arXiv},
  eprint       = {2311.07948},
}

@inproceedings{invariants4,
  author       = {Muhammad A. A. Pirzada and
                  Giles Reger and
                  Ahmed Bhayat and
                  Lucas C. Cordeiro},
  title        = {{LLM-Generated Invariants for Bounded Model Checking Without Loop Unrolling}},
  booktitle    = {ASE},
  pages        = {1395--1407},
  publisher    = {{ACM}},
  year         = {2024},
  }

@article{postconditions,
  author       = {Madeline Endres and
                  Sarah Fakhoury and
                  Saikat Chakraborty and
                  Shuvendu K. Lahiri},
  title        = {{Can Large Language Models Transform Natural Language Intent into Formal
                  Method Postconditions?}},
  journal      = {Proc. {ACM} Softw. Eng.},
  volume       = {1},
  number       = {{FSE}},
  pages        = {1889--1912},
  year         = {2024},
}

@inproceedings{specgen,
  author       = {Lezhi Ma and
                  Shangqing Liu and
                  Yi Li and
                  Xiaofei Xie and
                  Lei Bu},
  title        = {SpecGen: Automated Generation of Formal Program Specifications via
                  Large Language Models},
  booktitle    = {ICSE},
  pages        = {16--28},
  publisher    = {{IEEE}},
  year         = {2025},
}

@inproceedings{lemur,
  author       = {Haoze Wu and
                  Clark W. Barrett and
                  Nina Narodytska},
  title        = {Lemur: Integrating Large Language Models in Automated Program Verification},
  booktitle    = {ICLR},
  publisher    = {OpenReview.net},
  year         = {2024},
}

@article{autoverus,
  author       = {Chenyuan Yang and
                  Xuheng Li and
                  Md Rakib Hossain Misu and
                  Jianan Yao and
                  Weidong Cui and
                  Yeyun Gong and
                  Chris Hawblitzel and
                  Shuvendu K. Lahiri and
                  Jacob R. Lorch and
                  Shuai Lu and
                  Fan Yang and
                  Ziqiao Zhou and
                  Shan Lu},
  title        = {AutoVerus: Automated Proof Generation for Rust Code},
  journal      = {Proc. {ACM} Program. Lang.},
  volume       = {9},
  number       = {{OOPSLA2}},
  pages        = {3454--3482},
  year         = {2025},
}

@article{dafnybench,
  author       = {Chloe Loughridge and
                  Qinyi Sun and
                  Seth Ahrenbach and
                  Federico Cassano and
                  Chuyue Sun and
                  Ying Sheng and
                  Anish Mudide and
                  Md Rakib Hossain Misu and
                  Nada Amin and
                  Max Tegmark},
  title        = {{DafnyBench: {A} Benchmark for Formal Software Verification}},
  journal      = {Trans. Mach. Learn. Res.},
  volume       = {2025},
  year         = {2025},
}

@article{laurel,
  author       = {Eric Mugnier and
                  Emmanuel Anaya Gonzalez and
                  Nadia Polikarpova and
                  Ranjit Jhala and
                  Yuanyuan Zhou},
  title        = {{Laurel: Unblocking Automated Verification with Large Language Models}},
  journal      = {Proc. {ACM} Program. Lang.},
  volume       = {9},
  number       = {{OOPSLA1}},
  pages        = {1519--1545},
  year         = {2025},
}

@inproceedings{clover,
  author       = {Chuyue Sun and
                  Ying Sheng and
                  Oded Padon and
                  Clark W. Barrett},
  title        = {{Clover: Closed-Loop Verifiable Code Generation}},
  booktitle    = {{AI} Verification},
  series       = {LNCS},
  pages        = {134--155},
  publisher    = {Springer},
  year         = {2024},
}

@inproceedings{baldur,
  author       = {Emily First and
                  Markus N. Rabe and
                  Talia Ringer and
                  Yuriy Brun},
  title        = {Baldur: Whole-Proof Generation and Repair with Large Language Models},
  booktitle    = {FSE},
  pages        = {1229--1241},
  publisher    = {{ACM}},
  year         = {2023},
}

@article{daisy,
  author       = {{\'{A}}lvaro F. Silva and
                  Alexandra Mendes and
                  Ruben Martins},
  title        = {Inferring multiple helper Dafny assertions with LLMs},
  journal      = {CoRR},
  volume       = {abs/2511.00125},
  year         = {2025},
  eprinttype   = {arXiv},
}

@inproceedings{autoformalization1,
  author       = {Yuhuai Wu and
                  Albert Qiaochu Jiang and
                  Wenda Li and
                  Markus N. Rabe and
                  Charles Staats and
                  Mateja Jamnik and
                  Christian Szegedy},
  title        = {{Autoformalization with Large Language Models}},
  booktitle    = {NeurIPS},
  year         = {2022},
}

@inproceedings{autoformalization2,
  author       = {Aditi Kabra and
                  Jonathan Laurent and
                  Sagar Bharadwaj and
                  Ruben Martins and
                  Stefan Mitsch and
                  Andr{\'{e}} Platzer},
  title        = {{Can Large Language Models Autoformalize Kinematics?}},
  booktitle    = {FMCAD},
  publisher    = {{TU} Wien Academic Press},
  year         = {2025},
}

@inproceedings{leandojo,
  author       = {Kaiyu Yang and
                  Aidan M. Swope and
                  Alex Gu and
                  Rahul Chalamala and
                  Peiyang Song and
                  Shixing Yu and
                  Saad Godil and
                  Ryan J. Prenger and
                  Animashree Anandkumar},
  title        = {{LeanDojo: Theorem Proving with Retrieval-Augmented Language Models}},
  booktitle    = {NeurIPS},
  year         = {2023},
}

@inproceedings{rocq,
  author       = {Christine Paulin{-}Mohring},
  title        = {{Introduction to the Coq Proof-Assistant for Practical Software Verification}},
  booktitle    = {Tools for Practical Software Verification, LASER},
  series       = {LNCS},
  pages        = {45--95},
  publisher    = {Springer},
  year         = {2011},
}

@inproceedings{dafny,
  author       = {K. Rustan M. Leino},
  title        = {{Dafny: An Automatic Program Verifier for Functional Correctness}},
  booktitle    = {LPAR},
  series       = {LNCS},
  pages        = {348--370},
  publisher    = {Springer},
  year         = {2010},
}

@inproceedings{lean,
  author       = {Leonardo de Moura and
                  Sebastian Ullrich},
  title        = {{The Lean 4 Theorem Prover and Programming Language}},
  booktitle    = {CADE},
  series       = {LNCS},
  pages        = {625--635},
  publisher    = {Springer},
  year         = {2021},
}

@article{alphacode,
  author       = {Yujia Li and
                  David H. Choi and
                  Junyoung Chung and
                  Nate Kushman and
                  Julian Schrittwieser and
                  R{\'{e}}mi Leblond and
                  Tom Eccles and
                  James Keeling and
                  Felix Gimeno and
                  Agustin Dal Lago and
                  Thomas Hubert and
                  Peter Choy and
                  Cyprien de Masson d'Autume and
                  Igor Babuschkin and
                  Xinyun Chen and
                  Po{-}Sen Huang and
                  Johannes Welbl and
                  Sven Gowal and
                  Alexey Cherepanov and
                  James Molloy and
                  Daniel J. Mankowitz and
                  Esme Sutherland Robson and
                  Pushmeet Kohli and
                  Nando de Freitas and
                  Koray Kavukcuoglu and
                  Oriol Vinyals},
  title        = {{Competition-Level Code Generation with AlphaCode}},
  journal      = {CoRR},
  volume       = {abs/2203.07814},
  year         = {2022},
  eprinttype   = {arXiv},
  eprint       = {2203.07814},
}

@article{mbpp,
  author       = {Jacob Austin and
                  Augustus Odena and
                  Maxwell I. Nye and
                  Maarten Bosma and
                  Henryk Michalewski and
                  David Dohan and
                  Ellen Jiang and
                  Carrie J. Cai and
                  Michael Terry and
                  Quoc V. Le and
                  Charles Sutton},
  title        = {{Program Synthesis with Large Language Models}},
  journal      = {CoRR},
  volume       = {abs/2108.07732},
  year         = {2021},
  eprinttype   = {arXiv},
  eprint       = {2108.07732}
}

@inproceedings{filliatre2013why3,
  title={{Why3—where programs meet provers}},
  author={Filli{\^a}tre, Jean-Christophe and Paskevich, Andrei},
  booktitle=esop,
  pages={125--128},
  year={2013},
  organization={Springer}
}

\end{document}